\documentclass[preprint,showpacs,preprintnumbers,amsmath,amssymb,endfloats]{revtex4}
\usepackage{graphicx}
\usepackage{dcolumn}
\usepackage{bm}
\usepackage{amssymb}
\usepackage{amsmath}
\usepackage{latexsym}
\usepackage{longtable}
\usepackage{ulem}

\begin{document}

\title{Disagreement between capture probabilities extracted from capture and
quasi-elastic backscattering excitation functions}
\author{V.V.Sargsyan$^{1,2}$, G.G.Adamian$^{1}$, N.V.Antonenko$^{1}$, and
R.P.S.Gomes$^3$}
\affiliation{$^{1}$Joint Institute for Nuclear Research, 141980 Dubna, Russia\\
$^{2}$International Center for Advanced Studies, Yerevan State University,
0025 Yerevan, Armenia\\
$^{3}$Instituto de Fisica, Universidade Federal Fluminense, Av. Litor\^anea,
s/n, Niter\'oi, R.J. 24210-340, Brazil}

\date{\today }

\begin{abstract}
Experimental quasi-elastic backscattering and capture (fusion) excitation
functions are usually used to extract the $s$-wave capture  
probabilities for the heavy-ion reactions. We investigated the $^{16}$O+$%
^{120}$Sn,$^{144}$Sm,$^{208}$Pb systems at energies near and below the
corresponding Coulomb barriers and concluded that the probabilities
extracted from quasi-elastic data are much larger than the ones extracted
from fusion
 excitation functions at sub and deep-sub barrier energies. This
seems to be a reasonable explanation for the known disagreement observed in
literature for the nuclear potential diffuseness derived from both methods.
\end{abstract}


 \maketitle

\section{Introduction}
In the investigation of reaction mechanisms between heavy ions, it is very
important to know the diffuseness parameter of the nuclear potential between
the colliding nuclei, since it affects the height and shape of the Coulomb
barrier, and consequently the cross sections of the reaction mechanisms,
particularly the capture or fusion process, for which this barrier has to be
overcome. There are two very widely used approaches to derive the nuclear
diffuseness parameter from experimental data. The first one is the use of fusion
data at near barrier energies. The second approach is to extract this
parameter from experimental  elastic or quasi-elastic backscattering data.
Both approaches should lead to the same value of this parameter. However,
presently we find in literature large discrepancies in the nuclear potential
diffuseness parameter extracted from the two mentioned analyzes as, for
example, was found by Mukherjee et al. \cite{Mukher},  Gasques et al. \cite{Gasques},
and Evers et al. \cite{Evers}. In this work
we investigate the reasons for such discrepancies.

For the systems investigated in the present work ($^{16}$O\ + $^{120}$Sn, $%
^{144}$Sm, and $^{208}$Pb), the fusion process exhausts the capture cross
section $\sigma _{cap}(E_{\mathrm{c.m.}})$, and, thus,   capture and fusion can be
considered as similar or even identical processes. Furthermore, for those
asymmetric and tightly bound systems, and at energies close or below the
Coulomb barrier, the deep inelastic and breakup processes can be neglected,
and consequently quasi-elastic process can be defined simply as the sum of
elastic, inelastic and transfer processes. So, we have chosen to deal with
very simple systems and conditions in the present work.

The present paper is organized as follows.
In Sec.~II, we present the methods used for the extraction of  the $s$-wave capture 
probabilities  from capture (fusion) and
quasi-elastic backscattering excitation functions.
The obtained results are given in Sec.~III.
We then summarize in Sec.~IV.

\section{Extraction methods}
\subsection{Capture probabilities from experimental capture excitation function}
We start with the capture cross section. The physical meaning of the first
derivative of the function $E_{\mathrm{c.m.}}\sigma _{cap}(E_{\mathrm{c.m.}%
}) $, in respect to the energy $E_{\mathrm{c.m.}}$, can be elucidated by
considering the penetration probabilities for different partial waves $J$.
One can approximate the $J$ dependence of the transmission probability $%
P_{cap}(E_{\mathrm{c.m.}},J)$, at a given $E_{\mathrm{c.m.}}$, by simply
shifting the energy, as it was done recently by Sargsyan et al. \cite%
{Sargsyan13b,Sargsyan13c}:
\begin{equation}
P_{cap}(E_{\mathrm{c.m.}},J)\approx P_{cap}(E_{\mathrm{c.m.}}-\frac{\hbar
^{2}\Lambda }{2\mu R_{b}^{2}}-\frac{\hbar ^{4}\Lambda ^{2}}{2\mu ^{3}\omega
_{b}^{2}R_{b}^{6}},J=0),
\end{equation}%
where $\Lambda =J(J+1)$, $R_{b}=R_{b}(J=0)$ is the position of the Coulomb
barrier at $J=0$, $\mu =m_{0}A_{1}A_{2}/(A_{1}+A_{2})$ is the reduced mass ($%
m_{0}$ is the nucleon mass), and $\omega _{b}$ is the curvature of the $s$%
-wave potential barrier. Here, we use the same procedure of Sargsyan et al.
\cite{Sargsyan13b,Sargsyan13c} for the expansion of the height $%
V(R_{b},J)=V_{b}(J)$ of the Coulomb barrier up to second order in $\Lambda $%
:
\begin{equation}
V_{b}(J)=V_{b}(J=0)+\frac{\hbar ^{2}\Lambda }{2\mu R_{b}^{2}}+\frac{\hbar
^{4}\Lambda ^{2}}{2\mu ^{3}\omega _{b}^{2}R_{b}^{6}}.
\end{equation}%
Now, if we use the formula for the capture cross section,
\begin{equation}
\sigma _{cap}(E_{\mathrm{c.m.}})=\pi \lambdabar
^{2}\sum_{J=0}^{J_{cr}}(2J+1)P_{cap}(E_{\mathrm{c.m.}}-\frac{\hbar
^{2}\Lambda }{2\mu R_{b}^{2}}-\frac{\hbar ^{4}\Lambda ^{2}}{2\mu ^{3}\omega
_{b}^{2}R_{b}^{6}},J=0),
\end{equation}%
convert the sum over the partial waves $J$ into an integral, and express $J$
by the variable $E=E_{\mathrm{c.m.}}-\frac{\hbar ^{2}\Lambda }{2\mu R_{b}^{2}%
}$, we obtain the following simple expression \cite{Sargsyan13b,Sargsyan13c}%
:
\begin{equation}
\sigma _{cap}(E_{\mathrm{c.m.}})=\frac{\pi R_{b}^{2}}{E_{\mathrm{c.m.}}}%
\int_{E_{\mathrm{c.m.}}-\frac{\hbar ^{2}\Lambda _{cr}}{2\mu R_{b}^{2}}}^{E_{%
\mathrm{c.m.}}}dEP_{cap}(E,J=0)[1-\frac{4(E_{\mathrm{c.m.}}-E)}{\mu \omega
_{b}^{2}R_{b}^{2}}],
\end{equation}%
where $\lambdabar ^{2}=\hbar ^{2}/(2\mu E_{\mathrm{c.m.}})$ is the reduced
de Broglie wavelength and $\Lambda _{cr}=J_{cr}(J_{cr}+1)$. For values $J$
larger than the critical angular momentum $J_{cr}$, the potential pocket in
the nucleus-nucleus interaction potential vanishes and the capture does not
occur. To calculate $J_{cr}$ and $R_{b}$, we use the nucleus-nucleus
interaction potential $V(R,J)$ of Ref.~\cite{Pot,Pot2}. For the nuclear part
of the nucleus-nucleus potential, the double-folding formalism with the
Skyrme-type density-dependent effective nucleon-nucleon interaction is used.
For the systems that we investigate in the present work, with $Z_{1}\times
Z_{2}<2000$, where $Z_{1,2}$ are the atomic numbers of interacting nuclei,
the critical angular momentum $J_{cr}$ is large (from 54 to 62), $P_{cap}(E_{%
\mathrm{c.m.}},J=0)\gg P_{cap}(E_{\mathrm{c.m.}}-\frac{\hbar ^{2}\Lambda
_{cr}}{2\mu R_{b}^{2}},J=0)$ for energies around and below the barrier, and
the factor $1-\frac{4(E_{\mathrm{c.m.}}-E)}{\mu \omega _{b}^{2}R_{b}^{2}}$
in Eq. (4) very weakly influences the results of the calculations at this
energy range \cite{Sargsyan13b}. Therefore, Eq. (4) can be approximated as
\begin{equation}
\sigma _{cap}(E_{\mathrm{c.m.}})=\frac{\pi R_{b}^{2}}{E_{\mathrm{c.m.}}}%
\int_{0}^{E_{\mathrm{c.m.}}}dEP_{cap}(E,J=0).
\end{equation}
Multiplying this equation by $E_{\mathrm{c.m.}}/(\pi R_{b}^{2})$ and
differentiating over $E_{\mathrm{c.m.}}$, one obtains
\begin{equation}
P_{cap}(E_{\mathrm{c.m.}},J=0)=\frac{1}{\pi R_{b}^{2}}\frac{d[E_{\mathrm{c.m.%
}}\sigma _{cap}(E_{\mathrm{c.m.}})]}{dE_{\mathrm{c.m.}}}.
\end{equation}
From Eq. (6) one can observe that $\frac{d[E_{\mathrm{c.m.}}\sigma _{cap}(E_{%
\mathrm{c.m.}})]}{dE_{\mathrm{c.m.}}}$ has a physical interpretation in terms
of the $s$-wave transmission in the entrance channel, and therefore the $s$%
-wave transmission probability can be extracted with a good accuracy from
the experimental capture cross sections $\sigma _{cap}(E_{\mathrm{c.m.}})$
at energies near and below the Coulomb barrier. There are other methods to
derive Eq. (6), as it was previously done by Balantekin et al. \cite%
{Bala}.

\subsection{Capture probabilities from  experimental quasi-elastic backscattering data}
Now lets consider the quasi-elastic scattering at backward angles. For
reactions involving only tightly bound nuclei at low energies, one can write
the direct relationship between capture and backward quasi-elastic
scattering probabilities as
\begin{equation}
P_{qe}(E_{\mathrm{c.m.}},J)+P_{cap}(E_{\mathrm{c.m.}},J)=1.
\end{equation}
This relation is due to the conservation of the reaction flux, since any
loss from the backward quasi-elastic scattering channel contributes directly
to the capture and vise-versa \cite{Sargsyan13b,Sargsyan13c,Canto06}. For
experimentalists, usually it is much easier and simpler to measure
quasi-elastic scattering than capture (fusion). By this reason, using Eq.
(7) one finds  the relation
\begin{equation}
P_{cap}(E_{\mathrm{c.m.}},J=0)=1-d\sigma _{qe}(E_{\mathrm{c.m.}})/d\sigma
_{Ru}(E_{\mathrm{c.m.}})
\end{equation}
which is well suited for the extraction
of the $s$-wave capture probability  from
the experimental quasi-elastic backscattering probability $d\sigma
_{qe}/d\sigma _{Ru}$.
In Eq. (8) the $P_{qe}(E_{\mathrm{c.m.}},J=0)=d\sigma_{qe}/d\sigma _{Ru}$
was assumed to be the ratio of the
quasi-elastic scattering differential cross section and Rutherford
differential cross section at 180 degrees \cite{Timmers}. However,
experimentally it is not possible to take quasi-elastic data at 180 degrees,
but rather at backward angles in the range from 150 to 170 degrees. So, the
corresponding center of mass energies have to be corrected by the
centrifugal potential at the experimental angle \cite{Timmers}.

\section{Results of calculations}
From Eqs. (6) and (8) one observes that the $P_{cap}(E_{\mathrm{c.m.}%
},J=0)$ could be extracted either from the experimental capture or fusion
[Eq. (6)] or from quasi-elastic backscattering [Eq. (8)] excitation
functions. One could also say that the proposed methods of extracting the $s$%
-wave transmission probabilities are almost model-independent. In the
following we show the results obtained by both methods, for the three
systems under investigation.

In Fig.~1 we show the results for the $^{16}$O + $^{120}$Sn system. One can
see a good agreement between the extracted probabilities for the $^{16}$O + $%
^{120}$Sn reaction by both methods [Eqs. (6) and (8)] at energies near the
Coulomb barrier, but there is disagreement at deep sub-barrier energies. In
Fig.~2   we show the results for the $^{16}$O + $^{208}$Pb system. Now, the
deviations are dramatic, even at near barrier energies, and they increase
with decreasing energy under the barrier. Figure~3 shows the results for the
$^{16}$O + $^{144}$Sm system. Again, one can observe disagreement even at
energies not too much below the barrier. The capture probabilities (closed squares)
extracted from the experimental capture data
\protect\cite{SmCap2}  were shifted by 1 MeV to the lower energies,
in order to try to understand the
mismatching between the probabilities extracted from the experimental
quasi-elastic backscattering and capture (fusion) data. Indeed, one finds an
improvement at near barrier energies, but the disagreement is still
important at deep sub-barrier energies.

So, clearly there is a mismatch between quasi-elastic backscattering and
fusion (capture) experimental data. The explanation of this disagreement is,
therefore, required, either experimentally or theoretically. One consequence
of the overestimation of the capture probability at deep sub-barrier
energies when one uses quasi-elastic backscattering data, in comparison with
those when one uses fusion cross section data, is that the nuclear potential
diffuseness parameter extracted from quasi-elastic scattering data should be
larger than that extracted from the fusion excitation function data. This
is, indeed, what has been reported in literature \cite{Gasques,Evers}.
Since the theoretical predictions agree with the experimental
capture (fusion) cross sections for these systems (for example,
see Ref. \cite{Pot2}), we might suspect that a possible reason for the overestimation of
the capture  probability from the experimental quasi-elastic data at
deep sub-barrier energies is the underestimation of the total reaction
differential cross section, that is taken as the Rutherford differential
cross section at this energy regime.

\section{Conclusions}
We have found an overestimation of the $s$-wave capture
probability, at very low energies, extracted from the experimental
quasi-elastic backscattering data with respect to that extracted from the
experimental capture (fusion) excitation function. Then, it is not
surprising that there are reported large discrepancies in the nuclear
potential diffuseness parameters extracted from the analyzes of the
experimental quasi-elastic (or elastic) backscattering and capture (fusion)
data. We suggest that it is desirable to have experimental efforts to
measure precisely quasi-elastic backscattering excitation functions, as well
as further theoretical investigation on this subject.
Our study may be useful for current
experimental activities in the field, as it puts together different processes.

\vspace{1.cm}

P.R.S.G. acknowledges the partial financial support from CNPq and FAPERJ.
This work was supported by DFG, NSFC, and RFBR. The
IN2P3(France)-JINR(Dubna) Cooperation Programme is gratefully acknowledged.%
\newline

\newpage

\begin{figure}
\includegraphics[scale=1]{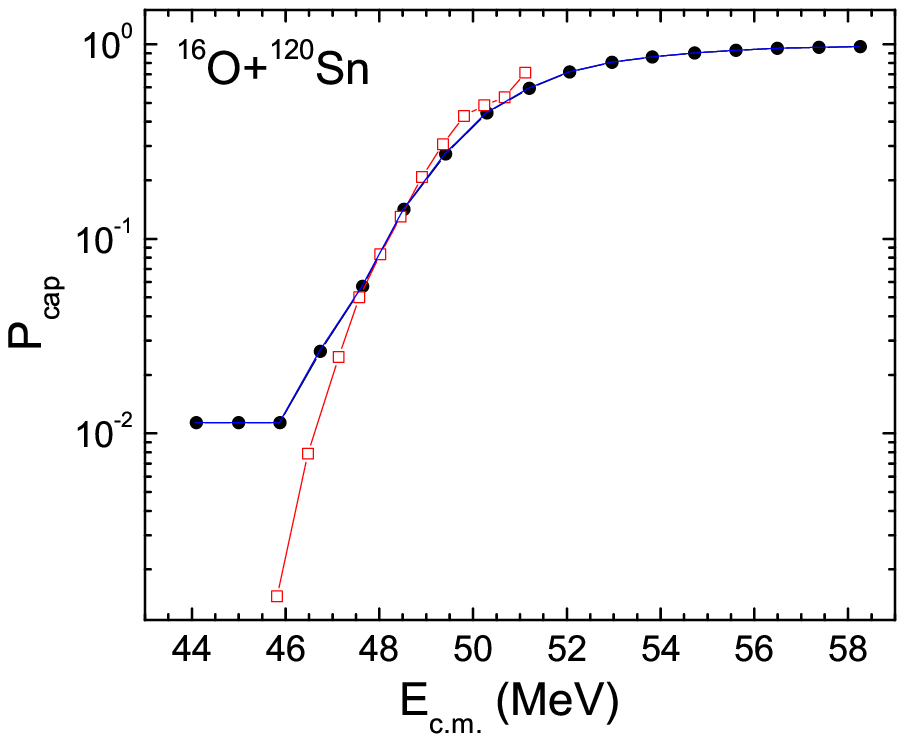}
\caption{(Color online)
The extracted   $s$-wave  capture  probabilities  for the     $^{16}$O + $^{120}$Sn  reaction
by employing  Eqs.~(6)  [squares]   and (8)  [circles].
The used experimental quasi-elastic backscattering and capture (fusion)
excitation functions are from Ref.~\protect\cite{Sinha}.
}
\label{1_fig}
\end{figure}

\begin{figure}
\includegraphics[scale=1]{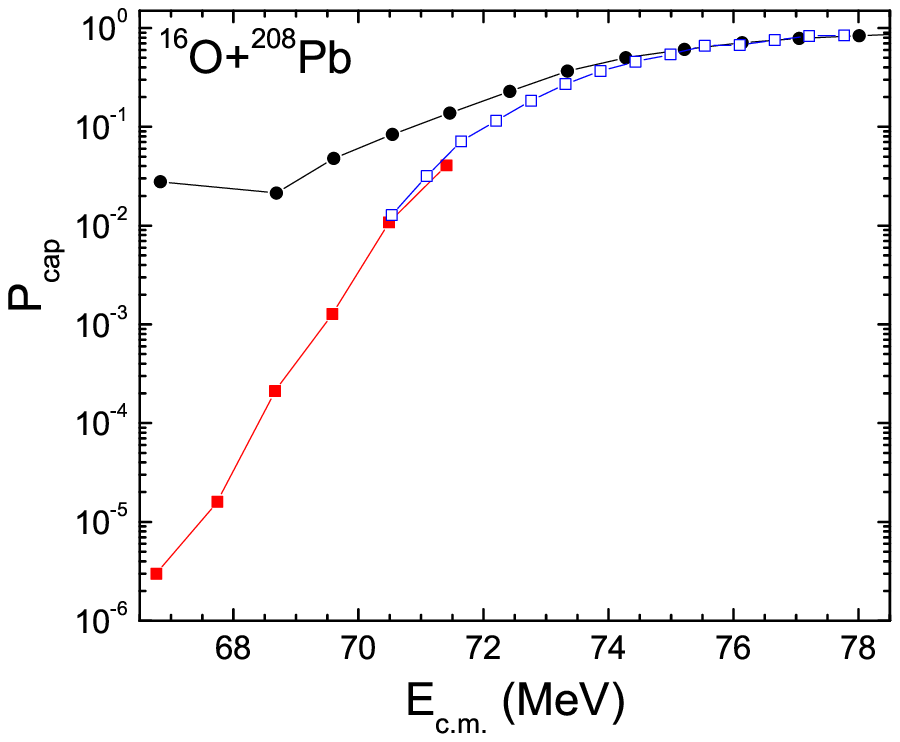}
\caption{(Color online) The extracted    $s$-wave  capture  probabilities  for the
$^{16}$O + $^{208}$Pb reaction  by employing
  Eqs.~(6)  [squares]   and (8)  [circles].
The used  experimental quasi-elastic backscattering
data are from Refs.~\protect\cite{Timmers2}.
The used experimental capture (fusion) excitation functions are from
Refs.~\protect\cite{Pbcap} (open squares)
and~\protect\cite{Pbcap3} (closed squares).
}
\label{2_fig}
\end{figure}

\begin{figure}
\includegraphics[scale=1]{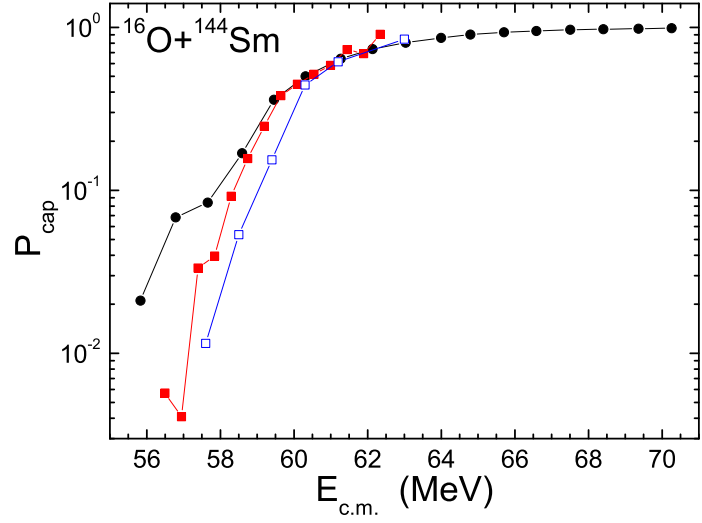}
  \caption{(Color online) The extracted  $s$-wave   capture probabilities for the
 $^{16}$O + $^{144}$Sm reaction by employing
   Eqs.~(6) [squares]   and (8) [circles].
 The
 experimental quasi-elastic backscattering and capture (fusion)  excitation functions
from Refs.~\protect\cite{Timmers} and~\protect\cite{SmCap1,SmCap2}, respectively.
The capture probabilities (closed squares)  extracted from the experimental capture (fusion)  excitation function
\protect\cite{SmCap2}  are shifted by 1 MeV to the lower energies (see text).}
  \label{3_fig}
\end{figure}

\end{document}